\documentclass{ws-tpe}
\usepackage{multicol}

\begin{document}

\catchline{1}{1}{2019}{}{}
\markboth{M. S. Feser \& I. Krumphals}{Viscous behavior of fluids in the eyes of adults: A global survey}

\title{Viscous behavior of fluids in the eyes of adults: A global survey}

\author{Markus Sebastian Feser}

\address{Physics Education, Universität Hamburg\\
	Hamburg, Free and Hanseatic City of Hamburg, Germany\\
	\email{markus.sebastian.feser@uni-hamburg.de}}

\author{Ingrid Krumphals}
\address{Institute for Secondary Teacher Education, University College of Teacher Education Styria\\
	Graz, Styria, Austria\\
	\email{ingrid.krumphals@phst.at}}

\maketitle


\begin{abstract}
The viscous behavior of fluids can be observed in numerous everyday situations. Therefore, it is reasonable to assume that people, since they are usually not specialists in fluids’ flow behavior, possess naïve conceptions about the viscous behavior of fluids. These conceptions more or less deviate from corresponding scientific explanations. Qualitative studies with preschool children and secondary school students from Germany have already identified various naïve conceptions about the viscous behavior of fluids (e.g., that the density or stickiness of a substance explains its viscous behavior). Within the present study, we explore the question of whether similar naïve conceptions can also be found among adults around the globe. To this end, based on previous research, an online questionnaire was developed to survey adults worldwide regarding their naïve conceptions about the viscous behavior of fluids. The survey was conducted anonymously, online, and voluntarily in spring 2023; participants were recruited via  {\it SurveySwap}. A total of 406 adults from all regions of the world (primarily Europe and North America) participated in the survey. In this paper, we report and discuss the main findings of this online survey. 
\end{abstract}


\begin{multicols}{2}

\section{Theoretical background and aims of the study}

Viscosity is an intensive property of fluids and describes their resistance to deformation when subjected to force. At the macroscopic level, other intensive properties—for example, density, pressure, and temperature—influence a fluid’s viscosity.\cite{[1],[2]} Nevertheless, the main causes of a fluid’s viscous behavior can be found at the sub-microscopic level.\cite{[1]} Primarily, a fluid’s viscosity is determined by the strength of the intermolecular attractions between its constituent particles,\cite{[3]} as well as the shape and size of these particles,\cite{[4]} because both influence the degree to which the constituent particles of a fluid can move past each other when the fluid is deformed. 

Since every fluid is more or less viscous, the viscous flow behavior of fluids can be observed in various everyday situations, for example, when spooning honey out of a jar, when squeezing shampoo from a bottle, or when spreading wall paint on a wall. For this reason, it is quite plausible to assume that people, since they are usually not specialists in fluids’ viscous behavior, possess naïve conceptions\footnote{Within the present paper, we use the term {\it naïve conceptions} to address underlying patterns in individuals’ explanations of scientific facts and phenomena that often stem from everyday life experiences. In this regard, the terms {\it students’ conceptions} or {\it children’s ideas} are generally more common within science education research.\cite{[5]} However, since the present study focuses on adults who are not necessarily students or learners, these two terms are improper for the present study, which is why we instead opted for the term {\it naïve conceptions}.} to explain such phenomena of everyday life, which more or less deviate from scientific explanations for these phenomena.\cite{[6],[7]} However, instead of naïve conceptions about fluids' viscosity, a clear understanding of it is of great benefit, even for one's day-to-day life. For example, as Paulun et al. put it, “[t]he ability to perceive viscosity is not only useful in its own right—for example, when judging whether milk has gone off, or whether eggs are sufficiently beaten—but also presumably reflects more general perceptual abilities to recognize objects, textures and materials that have highly mutable appearance.” (Ref.~\refcite{[19]}, p. 163). Nevertheless, the viscous behavior of fluids represents a rather subordinary topic within the curricula for school science or for natural science courses within higher education.\cite{[8]} Presumably for this reason, to the best of our knowledge, only a handful of studies from physics education research have been conducted regarding individuals’ naïve conceptions about the viscous behavior of fluids so far.

In qualitative studies conducted with preschool children and secondary school students from Germany, several naïve conceptions about the viscous behavior of fluids have been identified.\cite{[9],[10]} Additionally, the results of the accompanying research from a professional development project addressing the development of diagnostic skills of preservice physics teachers imply that preservice physics teachers tend to have the same naïve conceptions about the viscous behavior of fluids as preschool children and secondary school students.\cite{[8]} Specifically, these studies revealed, among further naïve conceptions, that a remarkable number of participants link the viscous behavior of a fluid to its density or its stickiness. They either equate a fluid’s viscosity with its density or its stickiness, or they assume a strict and direct proportionality between viscosity and density and/or viscosity and stickiness. Both of these conceptions are erroneous from a scientific point of view. While viscosity describes a fluid’s resistance to deformation, density is a measure of its mass (or number of particles) per unit volume, and stickiness refers to the inherent tackiness or initial adhesion of a fluid (see also Refs.~\refcite{[2],[11]}). Furthermore, easily observable phenomena from everyday life illustrate that there is no strict and direct proportionality between viscosity and density or viscosity and stickiness. For example, sunflower oil floats on water due to its lower density, but it has a higher viscosity than water, and ketchup is a more viscous but less sticky fluid than honey.

Given that previous studies addressing individuals’ naïve conceptions about the viscous behavior of fluids were based on comparatively small samples of preschool children, secondary school students, and preservice physics teachers (6 $\leq$ N $\leq$ 27;\linebreak see Refs.~\refcite{[8],[9],[10]}) and were solely conducted in German-speaking countries (Austria and Germany), the transferability of their findings to individuals with different backgrounds and within other living contexts remains unexplored so far, especially their transferability to physics students at universities or more broadly to adults in general. In the present study, we therefore aim to complement previous research on individuals’ naïve conceptions about the viscous behavior of fluids in order to reduce this research gap to some degree. More precisely, we administered a global online survey among adults based on the following exploratory research questions:

\medskip

\noindent {\bf RQ1:} To what extent do adults around the globe reveal naïve conceptions about the viscous behavior of fluids that have been identified in previous qualitative research?

\medskip

\noindent {\bf RQ2:} To what extent does a global survey on adults’ naïve conceptions about the viscous behavior of fluids reveal:

\begin{alphlist}[(5)]
\item differences between participants with different demographic features (gender identification, nationality, educational background)?

\medskip

\item correlational relationships between participants’ naïve conceptions about the viscous behavior of fluids and their self-evaluated knowledge about natural science in general and about the viscous behavior of fluids in particular?
\end{alphlist}

Below, we describe the methodological approach of the present study and detail our surveyed sample. Subsequently, we report and discuss the results of our data analysis regarding research questions RQ1 and RQ2.

\begin{figure*}
	\begin{center}
		\includegraphics[width=0.95 \textwidth]{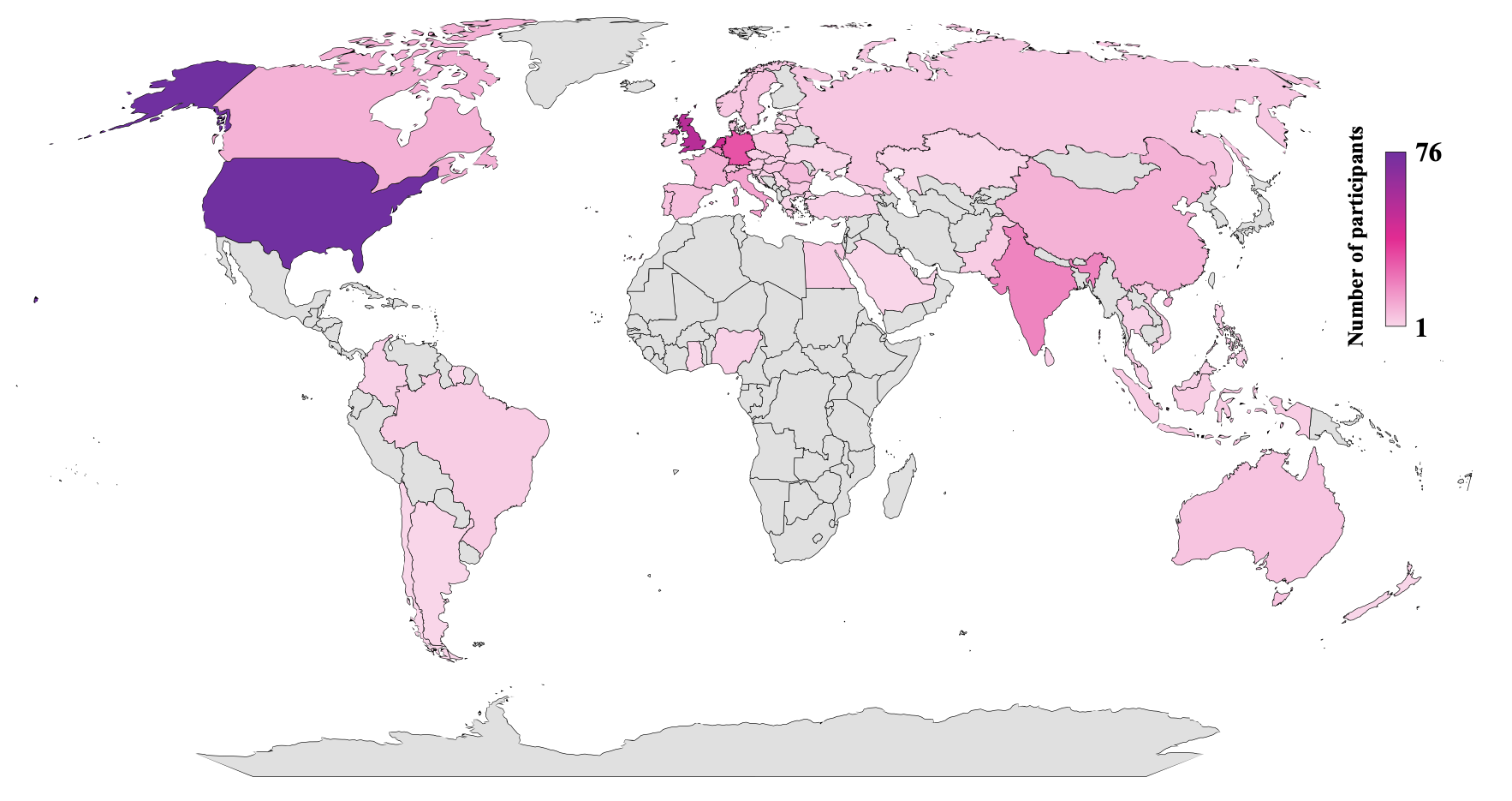}
	\end{center}
	\caption{Global distribution of all participants (N = 406).}
	\label{figure1}
\end{figure*}

\section{Method}

Drawing on our previous studies, we designed an online questionnaire which we used to survey a total of N = 406 adults around the globe (for details, see Sample). Data collection was carried out in line with the legal and ethical standards for educational research in Austria and Germany;\cite{[12]} all participants were surveyed voluntarily and anonymously and were informed beforehand about the aim of the present study. The survey was conducted from March to May 2023, and the participants were recruited via the online recruiting portal {\it SurveySwap} (\url{www.surveyswap.io}).

All questions in our online questionnaire were designed in a closed-response format and were administered to participants written in English. The order of the questions within the questionnaire was identical for all participants. In summary, our developed questionnaire includes the following instruments:

\begin{romanlist}[(5)]
	\item questions regarding participants’ demographic features (age, gender identification, nationality, educational background);
	
	\medskip
	
	\item participants’ self-evaluation of their own knowledge about natural science in general and about the viscous behavior of fluids in particular (response format: bipolar rating scale, ranging from 1—very poor to 10—excellent);
	
	\medskip
	
	\item two monitoring questions\footnote{These two montoring questions were “Have you participated in this survey before?” and “Please tick ‘true' to show that you have answered the questionnaire conscientiously.” These two questions were the first and last questions of our survey. Responses from participants who responded “yes” to the first question and/or “false” to the second question were excluded from the dataset.} to exclude “dishonest” responses to our questionnaire from the dataset (excluded “dishonest” responses: \linebreak N = 56; total sample size: N = 462); and
	
	\medskip
	
	\item an instrument we developed for assessing participants’ naïve conceptions about the viscous behavior of fluids that consists of 10 statements to which participants are asked to respond on a dichotomous true-false scale (see Table \ref{table2}). Each of these 10 statements represents a naïve conception about the viscous behavior of fluids that we identified in our previous qualitative research (see Refs.~\refcite{[8],[9],[10]}) and are all erroneous from a scientific point of view (e.g., “viscosity and density are the same thing” or “water has no viscosity”). Prior to the present study, we piloted and discursively adjusted this instrument with a doctoral student conducting physics education research to ensure it is bug-free and comprehensible for participants. Within the present study, the reliability of our developed instrument was examined based on the responses of our 406 participants and by calculating the greatest lower-bound estimate of reliability (glb coefficient)\footnote{Since a correlation analysis indicated that the 10 statements of our developed instrument are highly heterogeneous (mean inter-item correlation = 0.10) and that our instrument is multidimensional, we calculated the glb coefficient instead of Cronbach’s $\alpha$ to examine the reliability of our developed instrument (for details, see Ref.~\refcite{[13],[14],[15]}).}. Our instrument reached a coefficient of glb = 0.702. Therefore, our instrument’s reliability can be considered sufficient.
	
\end{romanlist}

\begin{table*} 
	\tbl{Descriptive statistics of the surveyed sample (N = 406).}
	{\begin{tabular}{@{}llccccc@{}} \toprule
			&  & & & \multicolumn{3}{c}{Score} \\
			\cline{5-7} \\[-8pt]
			Variable & Category & N & \% & M & SD & 95\% CI \\ \colrule
			Gender idenfication & Female & 245 & 60.3 & 3.8 & 2.0 & [3.6, 4.1] \\
			& Male & 149 & 36.7 & 3.6 & 2.0 & [3.3, 4.0]\\
			& Neither female nor male & 7 & 1.7 & 4.4 & 2.7 & [2.7, 8.5]\\
			& Prefer not to say & 5 & 1.2 & 3.6 & 0.9 & [2.8, 4.4]\\
			Geographical origin
(world region) & Africa & 6 & 1.5 & 4.0 & 2.1& [0.2, 5.4]\\
			& Asia & 54 & 13.3 & 4.3 & 2.1 & [3.7, 4.9] \\
			& Europe & 241 &	59.4 &	3.7	& 1.9 & [3.5, 3.9] \\
			& North America &  85 & 20.9 &	3.7 & 2.2 & [2.9, 3.9]\\
			& Oceania &  6	 & 1.5	& 4.7	& 2.0 & [3.1, 9.7]\\
			& South America &  8 &	2.0	& 3.8	& 1.6 &[2.6, 5.5]\\
			& Prefer not to say &  6 &	1.5 & 5.2 &	2.9 & [2.9, 8.9] \\
			Highest level of education & 12th grade or lower &  27 &	6.7	 & 3.9	& 2.2 & [3.1, 4.9]\\
			& High school degree or equivalent &  93 & 22.9 & 3.4	& 1.9 & [3.0, 3.8] \\
			& Bachelor’s degree or equivalent &  195 &	48.0 & 3.7 & 2.0 & [3.5, 4.0] \\
			& Master’s degree or PhD &  88 &	21.7 & 4.1 & 2.0 & [3.7, 4.5] \\
			& Prefer not to say &  3 &	0.7	 & 3.0	& 2.7 & [-4.6, 4.9] \\ \botrule
		\end{tabular} \label{table1}
	}
	\begin{tabnote}
		N = number of participants; \% = percentage frequency based on the total number of participants; Score = participants' score on our instrument for assessing naïve conceptions about the viscous behavior of fluids (mean, standard deviation, 95\% confidence interval calculated via bootstrapping).\\
	\end{tabnote}
\end{table*}

For addressing RQ1, we calculated the percentage frequencies in which participants rated the 10 statements of our instrument for assessing naïve conceptions about the viscous behavior of fluids as true or false. Additionally, using respective bootstrapping techniques, the corresponding 95\% confidence intervals to these percent frequencies were determined.\cite{[16],[16b]} To address RQ2b, we determined the bivariate correlation (Pearson’s r) between the participants’ score on our instrument for assessing naïve conceptions about the viscous behavior—which we defined as the number of statements a participant rated as true—and their self-evaluated knowledge about science in general and about the viscous behavior of fluids in particular.\footnote{Within this correlation analysis, 95\% confidence intervals were calculated via bootstrapping.} Prior to this analysis, we used a scatterplot to check the relationship between these three variables for linearity, and we used Q-Q plots to check whether they were normally distributed.\cite{[17]} Finally, to address RQ2a, we conducted a total of three one-way analyses of variance in which the participants’ score on our instrument for assessing naïve conceptions about the viscous behavior was the dependent variable and their gender identification, geographical origin (determined based on their nationality), and their highest level of education were utilized as the independent categorical variables. Responses from participants who reported no information on one of these categorical variables (see Sample) were excluded from these data analyses. Furthermore, given that some categories of these categorical variables contained only a small number of participants (e.g., our sample includes only six participants from Oceania and seven participants who identify as neither female nor male; see also Table \ref{table1}), we utilized Kruskal–Wallis–Tests as the nonparametric equivalent for the parametric one-way analyses of variance.\cite{[18]} 

\section{Sample}
Our 406 participants had an average age of 27.8 years (SD = 7.1 years) by the time we surveyed them and originated from 60 different countries around the globe (see Figure \ref{figure1}). 

Table \ref{table1} summarizes the key descriptive statistics of our surveyed sample. In terms of gender identification, the majority of our participants identified as female, constituting 60.3\% of our sample, while males accounted for 36.7\%. A small proportion, 1.7\%, indicated that they identify as neither female nor male, and 1.2\% preferred not to disclose their gender identification. Geographically, the largest segment of participants hailed from Europe, at 59.4\%, followed by North America, at 20.9\%. Asia accounted for 13.3\% of participants, while smaller percentages were attributed to Africa (1.5\%), Oceania (1.5\%), and South America (2.0\%). Notably, a fraction of participants (1.5\%) preferred not to specify their geographical origin. The highest level of education attained by the participants varied, with 48.0\% holding a bachelor’s degree or equivalent, and 21.7\% having completed a master’s degree or PhD. An additional 22.9\% possessed a high school diploma or equivalent, and 6.7\% had completed their education at or below the 12th-grade level. A very small percentage (0.7\%) opted not to disclose their highest level of education.

Beyond that, our participants self-evaluated their knowledge about natural science in general \linebreak (M = 5.28; SD = 1.93; 95\% CI [5.10, 5.47]) and their knowledge about the viscous behavior of fluids (M = 4.61; SD = 2.32; 95\% CI [4.38, 4.84]) both as moderate. However, participants’ mean self-evaluated knowledge about natural science in general was significantly higher than their mean self-evaluated knowledge about the viscous behavior of fluids (small effect size: d = 0.32; t(405) = 6.62; \linebreak p $<$ 0.001).

\begin{figure*}
	\begin{center}
		\includegraphics[height=0.91 \textheight]{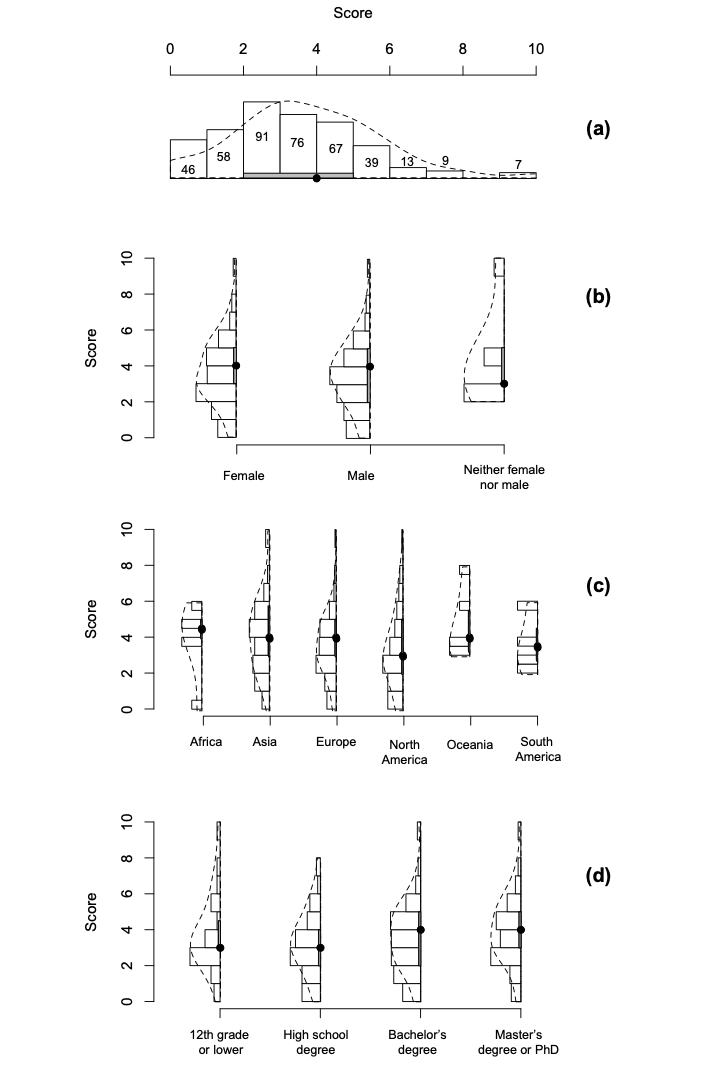}
	\end{center}
	\caption{Histograms, box plots, and density plots of participants' score on our instrument for assessing naïve conceptions about the viscous behavior of fluids: (a) total sample; (b) participants with different gender indentifications; (c) participants with different geographical origins; (d) participants with different highest levels of education.}
	\label{figure2}
\end{figure*}

\section{Results}

On average, our participants rated 3.74 (SD = 1.99; 95\% CI [3.55, 3.94]) of the 10 statements constituting our instrument for assessing naïve conceptions about the viscous behavior of fluids as true (for the mean scores and distributions of participants with different demographic features, see Table \ref{table1} and Figure \ref{figure2}). Given that these 10 statements are all erroneous from a scientific point of view (see Method) and only 5.7\% (N = 23) of our participants correctly rated all these statements as false, it is safe to conclude that virtually all our participants possess some degree of naïve conceptions about the viscous behavior of fluids, consistent with those conceptions found in qualitative research with preschool children and secondary school students. Additionally, given the fact that only 1.7\% (N = 7) of our participants rated all 10 statements of our instrument as true, it is reasonable to assume that potential agreement and/or straight-lining biases within the response behaviour of our participants are (if at all) marginal.

Table \ref{table2} summarizes the results of our data analysis regarding RQ1. To begin with, the response behavior of some participants revealed naïve conceptions regarding how temperature and pressure influence a fluid’s viscosity as 16.2\% \linebreak (95\% CI [12.8\%, 20.1\%]) of our participants inaccurately stated that fluids of the same temperature possess the same viscosity and 33.7\% \linebreak (95\% CI [29.2\%, 38.6\%]) stated that fluids of the same pressure have the same viscosity. Furthermore, another naïve conception emerged as 34.0\% \linebreak (95\% CI [29.4\%, 38.6\%]) of our participants erroneously asserted that water has no viscosity. However, the most compelling trend within the results of our data analysis lies in the participants’ entanglement of viscosity with other properties of fluids. Astonishingly, a noteworthy 19.0\% \linebreak (95\% CI [15.3\%, 23.0\%]) of our participants held the naïve conception that viscosity and density are interchangeable, and 22.9\% \linebreak (95\% CI [18.9\%, 27.3\%]) deemed viscosity and stickiness to be synonymous. Similarly, 67.0\% \linebreak (95\% CI [62.4\%, 71.5\%]) of our participants incorrectly associated viscosity with density, assuming that the more viscous a fluid is, the denser it is, and 45.8\% (95\% CI [40.9\%, 50.8\%]) exhibited misunderstandings about viscosity and stickiness, assuming a strict and direct proportionality between these two properties. Moreover, 35.2\% \linebreak (95\% CI [30.7\%, 40.1\%]) erroneously assumed that viscosity directly impacts a fluid’s volume, and 50.0\% (95\% CI [45.0\%, 55.0\%]) believed viscosity to be proportionally connected to the fluid’s mass or the matter it contains. 

In the Kruskal-Wallis tests (RQ2a), no statistically significant differences regarding participants’ score on our instrument for assessing naïve conceptions about the viscous behavior of fluids emerged between participants with different geographical origins (H(5) = 8.87, p = 0.114), participants identifying with different genders \linebreak (H(2) = 0.65, p = 0.722), or participants with different highest levels of education (H(3) = 5.41, \linebreak p = 0.144). In line with that, the 95\% confidence intervals listed in Table \ref{table1} also indicate that there are no significant differences between any subgroups (for an illustration, see also the histograms, box plots, and density plots in Figure \ref{figure2}). However, in line with our expectations,\footnote{Given that a high score on our instrument indicates that participants consider a large number of inaccurate statements about the viscous behavior of fluids to be correct, we expected negative correlations as the outcome of our data analysis for RQ2b.}  our data analysis for RQ2b revealed weak negative but significant correlations between the participants’ score on our instrument for assessing naïve conceptions about the viscous behavior of fluids and their self-evaluated knowledge about natural science in general (r = –0.14; \linebreak p = 0.005; 95\% CI [-.24, -.04]) and between the participants’ score and their self-evaluated knowledge about the viscous behavior of fluids (r = –0.20; \linebreak p $<$ 0.001; 95\% CI [-.29, -.09]).

\begin{table*}
	\tbl{Participants’ responses to our instrument for assessing naïve conceptions about the viscous behavior of fluids.}
	{\begin{tabular}{@{}lcccc@{}} \toprule
			&  & & \multicolumn{2}{c}{95\% confidence interval} \\
			\cline{4-5} \\[-8pt]
			Statement & $\text{\%}_\text{true}$ & $\text{\%}_\text{false}$ & $\text{\%}_\text{true}$ 95\% CI & $\text{\%}_\text{false}$ 95\% CI \\ \colrule
			The more viscous a fluid is, the more dense it is. &	67.0\% &	33.0\% &	[62.4\%, 71.5\%]	& [28.5\%, 37.6\%]  \\
			The more viscous a fluid is, the more sticky it is. &	45.8\% &	54.2\% &	[40.9\%, 50.8\%] &	[49.2\%, 59.1\%]  \\
			The more viscous a fluid is, the greater its volume is.	& 35.2\%	& 64.8\% &	[30.7\%, 40.1\%] &	[59.9\%, 69.3\%]  \\
			The more viscous a fluid is, the greater its mass is. &	50.0\% &	50.0\%	 & [45.0\%, 55.0\%] &	[45.0\%, 55.0\%]  \\
			The more viscous a fluid is, the more matter it contains. &	50.0\% &	50.0\% &	[45.0\%, 55.0\%] &	[45.0\%, 55.0\%]  \\
			Viscosity and density are the same thing. &	19.0\% & 81.0\%	& [15.3\%, 23.0\%] &	[77.0\%, 84.7\%]  \\
			Viscosity and stickiness are the same thing. &	22.9\%	& 77.1\% &	[18.9\%, 27.3\%] &	[72.9\%, 81.1\%]  \\
			Water has no viscosity.	& 34.0\% &	66.0\% & [29.4\%, 38.6\%] &	[61.1\%, 70.6\%] \\
			Fluids of the same temperature have the same viscosity.	 & 16.2\%	& 83.7\%	& [12.8\%, 20.1\%] & [79.8\%, 87.2\%] \\
			Fluids of the same pressure have the same viscosity. &	33.7\% &	66.3\% & [29.2\%, 38.6\%] &	[61.4\%, 70.8\%] \\ \botrule
		\end{tabular} \label{table2}
	}
	\begin{tabnote}
		Statement = Statement to which participants are asked to respond on a dichotomous true-false scale, $\text{\%}_\text{true}$ =  percentage frequency of participants that rated a statement as “true”, $\text{\%}_\text{true}$ =  percentage frequency of participants that rated a statement as “false”, $\text{\%}_\text{true}$ 95\% CI = 95\% confidence interval of $\text{\%}_\text{true}$ calculated via bootstrapping, $\text{\%}_\text{false}$ 95\% CI = 95\% confidence interval of $\text{\%}_\text{false}$ calculated via bootstrapping.\\
	\end{tabnote}
\end{table*}

\section{Limitations}

There are some limitations of the present study that need to be considered. First, since the order of the questions within our questionnaire was identical for all participants (see Method), the results of the present study may be biased by an ordering effect. Second, the way we recruited the participants of our study could introduce a self-selection bias, as those who volunteered to participate might also have a greater interest in natural sciences or a different level of knowledge about natural science compared to the general population. Third, the design of our instrument for assessing naïve conceptions about the viscous behavior of fluids might have influenced participants’ response behavior, as they might have considered it unlikely that all 10 statements of the instrument are erroneous from a scientific point of view. Therefore, it might be possible that different results will emerge if some of these statements are inverted into scientifically correct statements. Fourth, given that our online questionnaire was administered to participants written in English, participants with limited proficiency in English might have misunderstood some of our questions, leading to inaccuracies in their responses. Fifth and finally, compared to the world’s population, participants from Western regions of the world are substantially overrepresented in this study, which limits the generalizability of our results.

\section{Discussion}

The present paper presented the results of an online survey in which we explored the extent to which a total of 406 adults from around the globe reveal naïve conceptions about the viscous behavior of fluids that have been identified in previous qualitative studies with preschool children and secondary school students from Germany. Moreover, we investigated whether and to what extent our online survey reveals differences between participants with different demographic features (gender identification, nationality, highest level of education) as well as correlational relationships between the participants’ naïve conceptions about the viscous behavior of fluids and their self-evaluated knowledge about natural science in general and about the viscous behavior of fluids in particular.

Our online survey included 10 scientifically erroneous statements regarding the viscosity of fluids, which participants were asked to rate as true or false according to their perception. In line with the results of previous studies addressing individuals’ naïve conceptions about the viscous behavior of fluids,\cite{[8],[9],[10]} our participants’ response behavior indicates a considerable lack of scientific understanding about the viscous behavior of fluids, as they, on average, rated a notable portion of these statements as true despite their scientific inaccuracy. Only a small fraction (5.7\%) rated all these statements as false. The naïve conceptions revealed in the participants’ response behavior include erroneous perceptions about the impact of temperature and pressure on viscosity, misunderstandings about the viscosity of water, and confusions between viscosity, density, and stickiness. Notably, a substantial proportion of our participants equated a fluid’s viscosity with its density (19.0\%) or its stickiness (22.9\%).

Beyond that, within our data analysis, no statistically significant differences emerged between participants with different geographical origins, gender identifications, or highest levels of education regarding their scores on our instrument for assessing naïve conceptions about the viscous behavior of fluids, indicating that these naïve conceptions may be equally prevalent among adults around the globe regardless of their different backgrounds. However, since some of the demographic features were subjected to a rather coarse-grained analysis within the present study (e.g., participants’ geographical origin), a more fine-grained analysis in future research might detect differences that the present study did not uncover. Similarly, since only selected demographic features were recorded within the present study, the possibility of such differences regarding further demographic features cannot be ruled out and, therefore, may also be investigated in future research (e.g., differences between individuals pursuing/not pursuing a science-related occupation or individuals with/without a science-related higher education degree).

Finally, our data analysis revealed negative and significant correlations between the participants’ score on our instrument for assessing naïve conceptions about the viscous behavior of fluids and their self-evaluated knowledge about natural science in general and between the participants’ score and their self-evaluated knowledge about the viscous behavior of fluids in particular. On one hand, these results may be interpreted as follows: individuals who themselves evaluated their knowledge about natural science in general and/or about a fluid’s viscosity in particular as low may also be more likely to hold naïve conceptions about the viscous behavior of fluids. On the other hand, since both these correlations are relatively weak (–0.20 $\leq$ r $\leq$ –0.14), they might also indicate that many people may be unaware that they hold naïve conceptions about the viscous behavior of fluids to some extent and, therefore, misjudge their own knowledge about natural science and about a fluid’s viscosity. Hence, future research should clarify whether and to what extent these possible interpretations of our study’s correlative results may be supported by further evidence.

In conclusion, the results of the present study suggest that numerous adults around the globe hold naïve conceptions about the viscous behavior of fluids. Though the viscous behavior of fluids is evident in many everyday life situations, it is a hardly represented topic within school science curricula or curricula for natural science courses within higher education (see Theoretical background); our findings, therefore, are not surprising. The results of the present study thus highlight a potential need for change in this regard. It would be highly desirable if more learning environments within school science curricula, higher education, and nonformal science education settings (e.g., science centers, scientific TV shows) could be developed and implemented that address naïve conceptions about a fluid’s viscosity (e.g., confusions between viscosity, density, and stickiness) and that enable individuals to develop a deeper and scientifically backed understanding of the viscous behavior of fluids (for a list of existing learning environments for higher education and/or school settings, see Ref.~\refcite{[10]}). Such efforts would contribute to foster people’s understanding about fluids and about matter and its interactions and, therefore, ultimately to their scientific literacy in general.

\end{multicols}

\paragraph{Dr. Markus Sebastian Feser} works as a postdoc at the Faculty of Education of the Universität Hamburg in Germany. Before working at the Universität Hamburg, he studied education sciences, physics and mathematics at the Julius-Maximilians-Universität Würzburg in Germany. In 2019, he completed his PhD in education studies with his thesis on the role of language in physics teachers’ everyday assessment practice. His main research focus is the professional development of (pre-service) science teachers. Among other topics, his current research focuses on the sense of belonging of students and student teachers to science, the conceptions of school students of the viscous behavior of fluids, and the role of language in teaching and learning physics.

\paragraph{Prof. Dr. Ingrid Krumphals} has been a professor of physics education at the University of Teacher Education Styria in Austria since December 2020. Previously, she was a post-doctoral research fellow at the University of Graz and there her research focus was the development of diagnostic competence of preservice physics teachers. Ingrid Krumphals earned her PhD at the Austrian Educational Competence Centre of Physics at the University of Vienna. She also has 5 years of experience as a high school teacher in physics and mathematics. Currently, her research focuses on content-specific teaching and learning processes in physics teacher education and training as well as in physics education at secondary level.

\end{document}